\documentclass[11pt,twoside]{article}
\def\kps{km s$^{-1}$}

\usepackage{asp2004}
\usepackage{epsf}
\usepackage{psfig}
\usepackage{lscape}
\usepackage{graphicx}

\begin{document}
\title{V838 Mon and M31-RV: The Stellar Populations Angle}   %%% Fill in title
\author{M. H. Siegel}   %%% Fill in author names
\affil{University of Texas - McDonald Observatory}    %%% Fill in author affiliations
\author{H. E. Bond}   %%% Fill in author names
\affil{Space Telescope Science Institute}    %%% Fill in author affiliations

\begin{abstract} %%% Abstract to run on from here.
Insight into the origin of unusual events like the eruption of V838 Mon can be
obtained from studies of the stellar populations from which they arise.
V838 Mon lies in an intriguing region of the Galaxy, toward the warped outer edge
of the disk, with significant contributions from the Galactic thick disk
and the recently discovered Monoceros tidal stream.  The initial distance
measures placed V838 Mon in a jumbled region of the Galaxy but the recent
shorter distances make it highly likely that V838 Mon was a thin disk star --
likely in a spiral arm -- consistent 
with the recent detection of a young cluster in the
vicinity.  We compare V838 Mon to M31-RV, 
a red variable that erupted in the
bulge of M31 in 1988 and had a peak luminosity and spectral evolution very similar
to V838 Mon.  Archival HST images show no nebulosity or unusual stars at
M31-RV's projected location.  Moreover, the only stellar population in the field
is a canonic old bulge population.  This indicates that whatever the
origin of the red novae, the mechanism is likely independent of age and 
progenitor mass.  In particular, the B3V star seen in V838 Mon
it not a necessary part of the eruption mechanism.
\end{abstract}

%%% MAIN BODY OF TEXT GOES HERE. CONSULT "INSTRUCTIONS FOR AUTHORS USING
%%% LATEX2E MARKUP", SECTIONS 2.3-2.6 FOR HELP WITH EQUATIONS, FIGURES,
%%% AND TABLES.

%\section{}   %%% Top level section head (remove "%" symbol)
%\subsection{}   %%% Second level section head (remove "%" symbol)
%\subsubsection{}   %%% Lowest level section head (remove "%" symbol)
%\section*{}	%%% Unnumbered top level section head (remove "%" symbol)
%\subsection*{}   %%% Unnumbered second level section head (remove "%" symbol)
\section{Introduction}

Understanding events like the eruption of V838 Monocerotis can be helped by
identifying the population of stars from which they arise.  Stellar populations
are usually distinct in their age, abundance and kinematics.  Quantifying these 
properties can yield
vital information on their component stars.  For example, the age of a stellar population
indicates the upper limit of main-sequence stellar mass.
Associating unusual stars with distinct stellar populations can be particularly useful 
in a case like V838 Mon where knowledge of the progenitor
and/or remnant of the red nova is extremely limited.

Our understanding of the evolution of horizontal branch stars is a classic
example of this method.  Because some horizontal branch stars can be associated with globular
clusters of unique age and abundance, this has allowed constraint on models of horizontal
branch evolution.

In this contribution, we examine the stellar populations associated with V838 Mon and a similar
nova M31-RV.  Our analysis provides important information on these unusual
events.

\section{Galactic Stellar Populations -- A Primer}

V838 Mon and M31-RV erupted within the two large spiral galaxies of the Local Group.
The stellar populations of the large spirals are extremely complex and overlap in
their properties.
However, they can be roughly divided into four distinct groups (for a comprehensive treatment
of this subject, see Majewski 1993 or Freeman \& Bland-Hawthorn 2002).  The 
dominant stellar populations of the Milky Way, and their counterparts in M31, are:

$\bullet$ The Galactic Bulge - the central spheroidal population of old (12 Gyr or
globular cluster age), metal-rich ($[Fe/H]\sim-0.3$) 
stars (McWilliam \& Rich 1994; Zoccali et al. 2003) formed in the early fast evolution of the Milky Way.
M31's counterpart is very similar (Davidge et al. 2005; Sarajedini \& Jablonka 2005) but 
may be far more extended.

$\bullet$ The Galactic Thin Disk -- a planar distribution which dominates the light of the Galaxy.  
Although disk are best known for their spectacular spiral arms of gas and young stars, their 
stellar mass is dominated by a broad distribution
of populations over a large range of age and abundance.  It is rotationally
supported, the Milky Way disk having a rotational velocity of 220 \kps at the solar radius
(Casertano et al. 1990).  
The density distribution is exponential in radius and height above the Galactic midplane 
(Siegel et al. 2002, hereafter S02).  The disk also show evidence of both warping and 
flaring (Momaney et al. 2006).  The counterpart in 
M31 is broadly similar and also warped (Innanen et al. 1982; Bellazzini et al. 2003).

$\bullet$ The Galactic Thick Disk -- a structure similar to the thin disk but intermediate
in age (8-10 Gyr), abundance ($[Fe/H]\sim-0.7$) and rotational velocity (Allende-Preito 
etla. 2006; Casertano et al.).  It is less massive 
but much more extended
than the thin disk (S02).  The origin of it is controversial.  It is unclear if M31 
has a population fitting this description (Sarajedini \& Van Duyne 2001).

$\bullet$ The Galactic Halo -- an extended diffuse spheroid of old metal-poor stars 
enveloping the Galaxy.
The halo may not have a smooth distribution but may be comprised
of streams of stars tidally stripped from dwarf galaxies during past mergers (S02 and references therein).  
M31's halo was thought to be metal-rich but there are indications that its halo is similar to the Milky Way
(Guhathakurta et al. 2006).  M31's halo is also known to have streams (Ibata et al. 2001).

The critical aspect of these stellar populations is that they are {\it distinct} in their
properties, particularly their spatial distribution.  Certain regions of the Milky Way and M31
are dominated by one population or another, making it possible to associate a star with
a parent population and reliably impute the parent's properties to it.

\section{V838 Mon's Parent Stellar Population}

Unfortunately, 
V838 Mon occupies a particularly confused part of the Galaxy.  The third quadrant is known
to have multiple spiral arms (Moitinho et al. 2006), a strong warp in the Galactic disk
(Momaney et al. 2006)
and a stream of halo stars from a disrupted dwarf galaxy (Newberg et al. 2002), 
which passes through
Monoceros before possibly blending with the disk in Canis Major (see, e.g., Martinez-Delgado
et al. 2005 and references therein)

There in no {\it a priori} reason to assume that V838 Mon belongs to any of these populations.
The standard assumption, based on its low Galactic latitude, is that it erupted
from a thin disk star (or stars).
But, as we will show below, there are reasons to be skeptical of this intuitive assessment.

Figure 1 shows a color-magnitude diagram of the V838 Mon field obtained with the McDonald 0.8m 
telescope.  The figure shows the typical "blue edge" of the field stars.  This edge usually
runs straight up and down, delineating the main-sequence turnoff of the disks and halo.  Redenning
along the line of sight, however, moves it redward with increasing magnitude.

\begin{figure}[!ht]
\plotfiddle{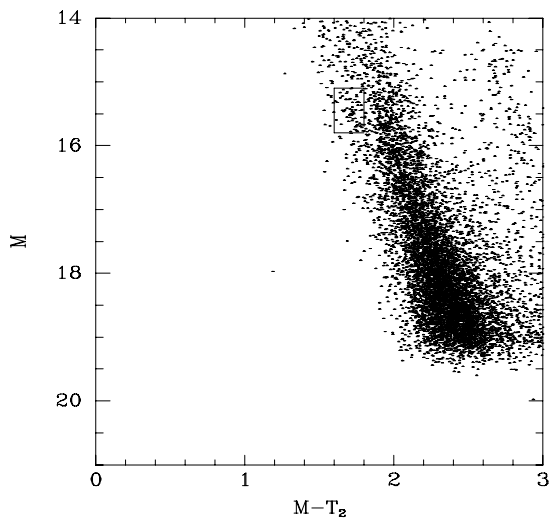}{2.5 in}{0}{90}{90}{-290}{-260}
\caption{Washington $MT_2$ photometry of the V838 Mon field.  $MT_2$ is roughly comparable to $VI$.  The
box highlights a statistical anomoly along the blue edge of the field stars -- a weak young MSTO that
traces either the star stream of the Monoceros dwarf galaxy or the Norma-Cygnus spiral arm.}
\end{figure}

The figure highlights a small bump just beyond the blue edge -- an apparently young main-sequence 
turnoff (detected primarily in a statistical test).  This "blue plume" has been used to trace the 
tidal stream of the merging dwarf galaxy
in the third quadrant (Bellazzini et al. 2004).  However, Carraro et al. (2006) have shown that 
the blue plume traces the 
Norma-Cygnus spiral arm better than it does the merging dwarf.  This would be a strong indicator
that V838 Mon's position on the sky is coincident with a spiral arm.

Kaminsky \& Pavlenko (2005) indicated an iron abundance for V838 of $[Fe/H]=-0.4$.  
This a bit rich for this region of the Galaxy (Munari et al. 2005) but within
the range of abundances known in either the thin or thick disk.

V838's position in dynamical phase space (velocity and position) could be a powerful discriminant.
We have constructed a simple Galactic model based on the density distributions detailed in S02 and
the velocity distributions of Casertano et al. (1990).
Figure 2 shows the range of radial velocities and distance above the Galactic stellar midplane 
dominated by the canonical stellar populations in this region.\footnote{The 
ordinate in figure 2 is Z', the distance from {\it stellar midplane}.  Z' takes into account
the Galactic warp as parameterized by Drimmel \& Spergel (2001) which reaches its maximum 
southward extent within 20 degrees of V838 Mon}.

The initial indications of a long distance to V838 Mon ($\sim$8 kpc - Tylenda 2004; Crause 
et al. 2005; Munari et al. 2005) were problematic for the inferred 
stellar populations.  Figure 2 shows the location of the nova in
$v_{helio}-Z'$ space (cross) assuming a distance of 8 kpc.  We can see that V838 Mon's 
classification is ambiguous.  The 
strong disk warp near V838 Mon would give the nova a Z' height of 770 pc.  The left panel of 
Figure 2 shows
that V838 Mon's radial velocity ($v_{helio}\sim65$ \kps - Claussen et al. 2005; 
Deguchi et al. 2005; Kipper et al. 2004) would be near the edge of both the thin
and thick disk distribution at this Z' while the right panel shows that about one-third of the 
stars
at this Z' are from the thick disk\footnote{The chance may be slightly lower
due to the presence of a thin disk flare in the third quadrant (Momaney et al.).  On the other
hand, as S02 show, the radial distribution of the disks is poorly constrained.}.  The 
Monoceros star stream is still only moderately constrained but comparison to the model
of Penarrubia et al. (2005) shows that V838 Mon's velocity and distance 
would be toward the edge of their tidal stream (see their figure 8).  The height
above the plane would be inconsistent with membership, however.

\begin{figure}[!ht]
\plotfiddle{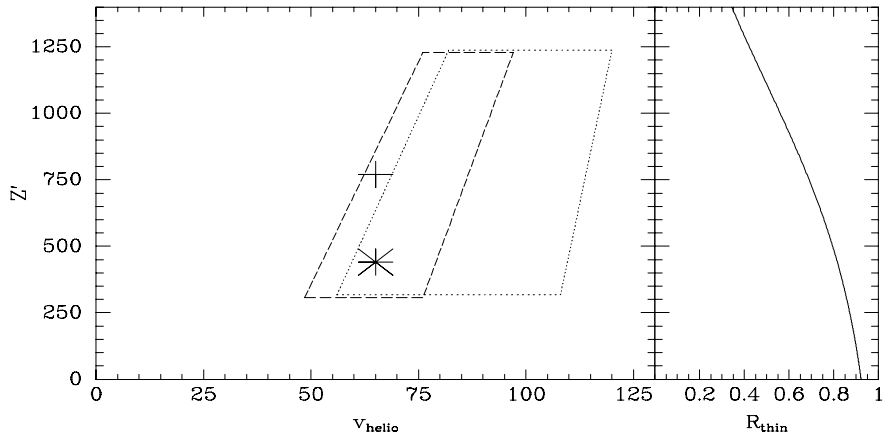}{2.75 in}{0}{90}{90}{-290}{-260}
\caption{V838 Mon's phase-space location.  In the left panel, the abscissa is the heliocentric
radial velocity.  In the right panel, it is the fraction of stars that are members of the thin
disk based on S02 model.  The ordinate is Z', the height above the stellar midplane, taking
into account the Galactic Warp.  Boxes denote the velocity distribution 
as function of Z' for the thin disk (dashed) and thick disk (dotted).  The cross denotes
the initial phase-space location of V838 based on a distance of 8 kpc.  The star represents
the phase-space location given a 6 kpc distance.}
\end{figure}

However, the shorter geometric distance of $\sim6$ kpc (Sparks, these proceedings) moves
V838 Mon to familiar territory (starred point on Figure 2).  It would
be in the middle of the thin disk velocity distribution
and its Z' would be even more dominated (80+\%) by the thin disk.  The new distance
would be also be too close to be a member of the Monoceros stream but could be
consistent with membership in a spiral arm.

Perhaps the final clinching evidence has been presented in these proceedings
by Asfar \& Bond, who show that V838 Mon appears associated not with just one B star
but with a small cluster of young blue stars.  If these stars are associated
with the nova, V838 could not possibly
be a member of Monoceros, the thick disk or the halo since none of those
populations has stars so young.

In summary, the parent stellar population of V838 Mon was far less clear before
this conference.  The latest results indicate that the intuitive conclusion about V838 Mon
is likely the correct one -- it is a part of a spiral arm and therefore arose from one of the
youngest populations in the Milky Way.

\section{M31-RV: A V838-Mon Clone?}

V838 Mon is not the first red nova to erupt in the local group.  An identical
event was observed in the bulge of M31 in 1988 (Rich et al. 1989, 
Tomaney \& Shafter 1992).  The event was brighter than $B\sim18.5$ (Sharov 1990, 1993; 
Boschi \& Munari 2004) for at least 80 days with maximum bolometric luminosity of $M_{bol}\sim-10$.
Unlike classical novae, M31-RV remained cool and red throughout
its evolution, with an initial spectrum of an M0 supergiant 
(Rich et al.) which gradually evolved to type
M5 and beyond (Mould et al. 1990; Boschi \& Munari).  It luminosity, color and evolution
are almost identical to that of V838 Mon.

The 1988 event was the only event in the last 50 years 
(Boschi \& Munari) -- a silence similar to  that seen
V838 Mon and V4332 Sgr (Goranskij, these proceedings.)

Little else is known about this rare event.  It was, of course, never targeted for observation
by HST because the telescope was launched two years later, long after the nova had faded.

In Bond \& Siegel (2006), we presented the result of a search of the HST archive for images of the
nova site.  Although M31-RV was never specifically designated for observation, the region
was serendipitously observed by WFPC2 in parallel mode during observations of M31's central
black hole.  The gem of these observations was a 1999 set of deep F555W-F814W images with the M31-RV
site positioned on the Planetary Camera.  There have also been recent observations with ACS that
have imaged the region in $B$.  Preliminary examination of these frames showed nothing unusual.

The stacked median-combined $VI$ images of the M31-RV site are shown in Figure 3.  We see no unusually
bright or unusually red stars in the region, no indication of nebulosity or a light echo, which
could have expanded to several arc-seconds over the eleven years since the nova.

\begin{figure}[!ht]
\plotfiddle{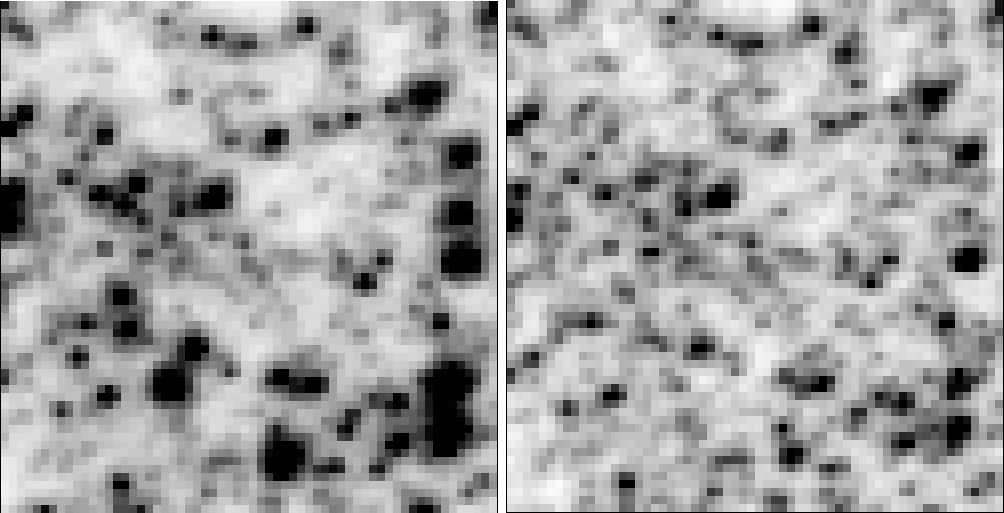}{3.75 in}{0}{45}{45}{-220}{20}
\caption{Stacked median images of a 3" by 3" area around the M31-RV eruption site.  Images are 
F814W (left) and F555W (right).}
\end{figure}

More insight is provided by the color-magnitude diagram
(Figure 4).  The diagram, which shows the entire field and a region within 3$\sigma$ of the
estimated M31-RV position, shows only a canonical old bulge population.  There is no young 
population; no spiral arm; no young cluster; no B3V star in the vicinity of M31-RV.
In short, the stellar population that produced M31-RV looks nothing like that which produced
V838 Mon.  M31-RV's remnant, if it is visible at all, is a typical bulge giant.

\begin{figure}[!ht]
\plotfiddle{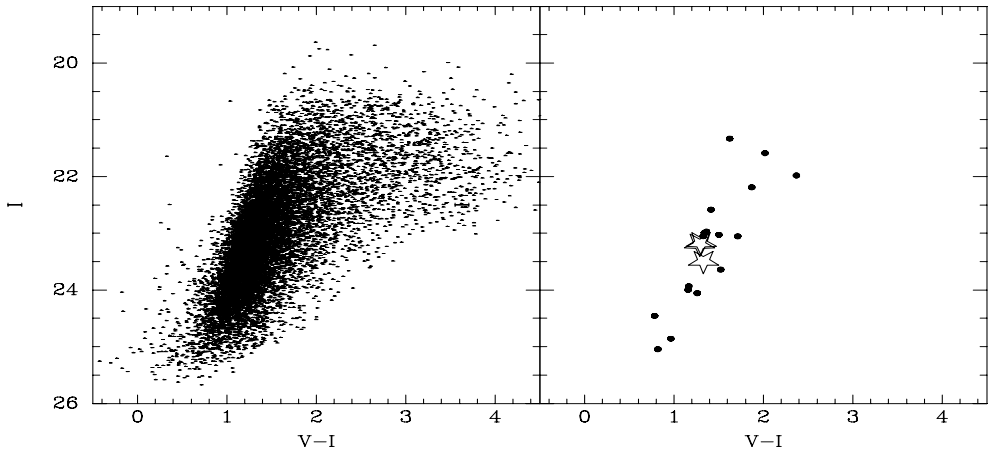}{3. in}{0}{90}{80}{-300}{-250}
\caption{$VI$ images of the region near M31-RV. Photometry of the entire region is shown to the
left.  The right panel shows stars within 1$\sigma$ (stars) and 3$\sigma$ (dots) 
of the projected M31-RV position.}
\end{figure}

\section{Discussion and Conclusions}

An analysis of the stellar populations associated with V838 Mon shows that it
likely erupted in a spiral arm from a young, moderately metal-rich population.  On
the other hand, M31-RV -- an event of identical brightness and evolution -- erupted in M31's
bulge from an old, metal-rich population.

That these identical events could occur in such diverse stellar populations indicates
that the mechanism producing the red variables is independent of age -- and stellar
mass.  Any mechanism that requires either an old evolved star or a young unevolved
star (such as the B3V star seen in V838 Mon) can not explain both events.

One encouraging note is that two -- or possibly three, if we include
V4332 Sgr -- red novae have erupted in the Local Group within the last twenty years.
Although there is no {\it a priori} reason to assume these events are isotropically
distributed in time, it gives hope that more will occur.  

Our analysis of the stellar populations shows that red variables can occur
anywhere in the Local Group.  A B3V star is not required.  High mass or evolved low-mass
stars are not required.  The most likely location of a future red novae is the disk
or bulge of the two large spiral galaxies, since these contain the bulk of the stellar 
populations.

Future synoptic surveys of the sky will provide far more information on 
the progenitors and pre-nova evolution of these rare events.  Our understanding of the
red novae -- whether they produce spectacular light echoes or not -- will vastly improve
with each new one that appears in the sky.

\acknowledgements %%% Text of acknowledgements runs on after this command.
MHS gratefully acknowledges the support of the American Astronomical Society and the 
National Science Foundation in the form of an International Travel Grant, 
which enabled him to attend this conference.  MHS also acknowledges financial
support from the conference organizers.

%%% THE BIBLIOGRAPHY
%%%
%%% CONSULT SECTION 4 OF "INSTRUCTIONS FOR AUTHORS" FOR HOW TO USE NATBIB.
%%% PLEASE USE THE "THEBIBLIOGRAPY" ENVIRONMENT
%%%

% edit and fill the lines below with the questions and answers, or 
% comment all lines except \end{document}

\question{Sugerman} How reliable is the HST position for M31-RV?

\answer{Siegel} We identified the position of M31-RV by directly registering
an old image of the RV and the stacked HST image to an intermediate deep
I-band 4-meter image.  So no assumptions are made about the reliability
of HST's pointing.

\question{Sugerman} Given the echo configuration, scattering angles suggest
you will not get much echo flux.  But since the field is so crowded, you
really have to do a difference-image analysis to establish if echoes are 
missing.

\answer{Siegel} I would certainly agree with that, especially since so many 
years have passed since the eruption event.

\question{Evans} We observed the area around M31RV with the Spitzer IRAC with 
the aim of getting a follow-up spectrum if we detected it - we didn't.

\answer{Siegel} That would indicate that M31-RV
has faded in the IR.  We'll have to see if V838 Mon has the same evolution
of its infrared luminosity.

\end{document}